\magnification=\magstep1
\hfuzz=6pt
\baselineskip=16pt

$ $
\vskip 1in
\centerline{\bf Quantum feedback with weak measurements}

\bigskip

\centerline{Seth Lloyd$^\dagger$}

\bigskip

\centerline {Jean-Jacques E. Slotine$^*$}

\bigskip

\centerline{\it$\dagger$ d'Arbeloff Laboratory for Information Systems
and Technology}

\centerline{\it$*$ Nonlinear Systems Laboratory}

\centerline{\it Department of Mechanical Engineering}

\centerline{\it Massachusetts Institute of Technology}

\vskip 1cm

\noindent{\it Abstract:} The problem of feedback control of
quantum systems by means of weak measurements is investigated
in detail.  When weak measurements are made on a set of
identical quantum systems, the single-system density matrix can be
determined to a high degree of accuracy while affecting each
system only slightly.  If this information is fed back into
the systems by coherent operations, the single-system density matrix can be
made to undergo an arbitrary nonlinear dynamics, including for example
a dynamics governed by a nonlinear Schr\"odinger equation. 
We investigate the implications of such nonlinear quantum
dynamics for various problems in quantum control and quantum 
information theory.   
The nonlinear dynamics induced by weak quantum feedback could
be used to create a novel form of quantum chaos in which the 
time evolution of the single-system wave function depends sensitively on 
initial conditions.

\vskip 1cm

The conventional theory of quantum feedback control 
assumes the use of strong or projective measurements
to acquire information about the quantum system under control (1-10).
Such measurements typically disturb the quantum system, destroying
quantum coherence and giving a stochastic character to quantum
feedback control.  But strong measurements are not the only
tool available for acquiring information about quantum systems (11-12).
In nuclear magnetic resonance (NMR), for example, one makes collective 
measurements on a set of effectively identical systems: by monitoring
the induction field produced by a large number of precessing
spins, one can obtain the average value of their magnetization
along a given axis while only slightly perturbing the states
of the individual spins (13).  We will call such measurements 
`weak measurements' since they only weakly perturb the individual
systems in the set. (Such weak measurements on large sets of 
identical systems should not be confused with
the {\it single-system} weak measurements debated in (14-16)).  
The information acquired by weak measurement
can then be fed back to the spins, for example to suppress super-radiant
decay (17-18).  NMR is not the only system in which weak measurement is 
possible: one can perform weak measurements on essentially any set of 
quantum systems that can be coupled weakly to an external apparatus. 
This paper provides a general theory
of quantum feedback control using weak measurements. 
Since weak measurements allow the accurate determination of the 
complete single-system density matrix of each member of a set
of identical quantum systems, while
effecting each system in the set arbitrarily weakly,
quantum feedback by weak measurement will be shown to be capable
of accomplishing tasks that are not possible using conventional,
strong measurements.  A model of quantum feedback using
weak measurements is given and applications are proposed.  In addition
to NMR, quantum feedback by weak measurements could be used in 
quantum optics and atomic and molecular systems to effect arbitrary 
nonlinear Schr\"odinger equations, to create solitons and
Schr\"odinger cats, to perform quantum computations,
and to institute novel forms of quantum chaos.

Quantum feedback via weak measurement represents a novel paradigm
for coherent control of quantum systems.  It allows the performance
of operations that are impossible in the normal, strong measurement
paradigm for quantum control.  For example, suppose that
each of the systems in the set is in the same unknown pure state.  Then
feedback with weak measurement can be used to drive them to any
desired pure state {\it reversibly}, while preserving quantum coherence.
This contrasts markedly to quantum feedback using strong measurements,
where a system in an unknown quantum state can be driven to any
desired quantum state, but only at the cost of disturbing the
system's state irreversibly and stochastically, destroying quantum
coherence in the process. 

The general picture of quantum feedback control using weak
measurements is as follows. 
Suppose that we have an set of $N$ identical non-interacting
quantum systems, each with density matrix $\rho$.  (Of course,
no set of quantum systems is perfectly non-interacting, but in
many situations --- e.g., liquid-state NMR, quantum optics ---
the non-interacting approximation holds to a high degree of accuracy.)
Assume that the system is coherently open-loop controllable,
so that we can perform arbitrary unitary transformations $U$
on the system (necessary and sufficient conditions for open-loop
coherent control of quantum systems are well known (1-6)). 
Now assume that we are able to make a sequence of collective
weak measurements on these systems
that allow us to determine the single-system reduced density matrix 
$\rho$ to some degree of accuracy
$\delta$, while disturbing this density matrix by an amount
$\epsilon$.  As will be seen below, both $\delta$ and $\epsilon$
can go to zero in the limit that the number of systems goes
to infinity.  
If the systems are individual nuclear spins, for example,
the single-spin density matrix can be determined by measuring the induction
signal produced about two different axes: this allows one to determine
the expectation of the magnetization along the $x$, $y$, and $z$ axes,
which is in turn sufficient information to determine the single-spin 
density matrix.  Now feed that information back into the set by
applying to each system a unitary transformation $U_\delta(\rho)$, 
where $U_\delta$ is some
potentially nonlinear function of $\rho$, and the subscript $\delta$
indicates that $U_\delta$ discriminates between different $\rho$ to an
accuracy $\delta$.  The time evolution of the system with feedback by 
weak measurement is accordingly given by
$$\rho'=U_\delta(\rho)(\rho+\Delta\rho)U^\dagger_\delta(\rho)\quad,
\eqno(1a)$$
where $ \Delta \rho$ is the perturbation to the single-system
reduced density matrix induced by the weak measurement, with
$\| \Delta \rho \| \leq \epsilon $ for a suitable norm $\| ~\|$ such
as the sup norm.
As will be shown below, in the limit $N\rightarrow\infty$, the collective
measurement can be performed in such a way that both 
$\delta$ and $\epsilon$ $\rightarrow 0$, and the time evolution 
of the single-system density matrix is governed by the equation
$$\rho'=U(\rho)\rho U^\dagger(\rho)\quad. \eqno(1b)$$.

The remainder of this paper will be devoted to exploring the 
implications of equations (1a) and (1b).  These equations have a variety of 
interesting features.  The first, perhaps most obvious, is that
they can be nonlinear as a function of $\rho$: 
if $\rho = \alpha\rho_1 + \beta \rho_2$,
it need not be the case that $U(\rho) \rho U^\dagger(\rho) =
\alpha U(\rho_1) \rho_1 U^\dagger(\rho_1) 
+\beta U(\rho_2) \rho_2 U^\dagger(\rho_2)$. 
(It is important to note that although the single-system
reduced density matrix undergoes a nonlinear evolution, the density
matrix for the set of systems taken collectively undergoes
a conventional linear time evolution: no laws of quantum
mechanics are broken in constructing this nonlinearity.)
If the weak measurement is made continuously
in time, then in the limit $N\rightarrow\infty$, $\delta\rightarrow 0$,
$\epsilon\rightarrow 0$, feedback causes the single-system density matrix
to obey a nonlinear Schr\"odinger equation 
$$\partial\rho/\partial t = -i[ H(\rho), \rho]\eqno(2)$$
\noindent where $H(\rho)$ is the Hamiltonian corresponding to 
$U(\rho)$.  Such nonlinearities in
the case of sets of nuclear spins are well-known: for 
example, if each nuclear spins in the set interacts with
the mean field generated by the spins taken together, then 
the single-spin density matrix obeys a nonlinear Bloch
equation (19).  Feedback by weak measurement allows one to impose
an {\it arbitrary} nonlinear Hamiltonian dynamics on the single-system
density matrices: if in 
the open-loop case, without feedback, one can apply any conventional
linear time evolution, then in the closed loop case, with feedback
of the results of weak measurements, one can apply any desired
nonlinear dynamics that preserves the eigenvalues of the density
matrix.  That is, one can take $\rho \rightarrow f(\rho)$, where
$f(\rho)$ has the same eigenvalues as $\rho$. If one can 
apply open-system operations (20-21) as well as closed-system, unitary 
transformations, then one can alter the
eigenvalues of the density matrix as well to take $\rho \rightarrow
g(\rho)$, where $g(\rho)$ can be an arbitrary density matrix.

Now let us look more closely at the dynamics of the weak
measurement process, in order to determine how accurately
the single-system density matrix can be measured and
at what cost.  There are two measures of the cost of
weak measurement:  first, the size $N$ of the set required
to attain a given accuracy $\delta$,  and second, the amount
$\epsilon$ by which the individual systems are perturbed by the 
weak interaction with the measuring apparatus.
Here we construct a specific model of weak measurement applicable 
to a wide range of physical systems. 

The general picture of measurement on $N$ identical systems
is as follows.  The density matrix for the systems is
$\rho_{\rm tot} = \rho\otimes\rho\otimes \ldots\otimes\rho$.
A positive operator valued measure (POVM) on this system
corresponds to a set of operators $\{A_{\mu i} \}$ such that
$\sum_{\mu i} A^\dagger_{\mu i} A_{\mu i} = {\bf I}$, where ${\bf I}$
is the identity operator; the measurement
corresponding to the POVM gives the result $\mu$ with probability $p_\mu =
{\rm tr} \sum_i A_{\mu i} \rho_{\rm tot} A^\dagger_{\mu i}$,
in which case the system is left in the state 
$\rho_{{\rm tot}\mu}=(1/p_\mu) \sum_i A_{\mu i} \rho_{\rm tot} 
A^\dagger_{\mu i}$ and the density matrix for the $\ell$'th subsystem
goes to $\rho_{\ell\mu}= {\rm tr}_{\ell'\neq \ell}
\rho_{{\rm tot}\mu}$.  

We will define a weak measurement to be
one that leaves the single system density matrices
unchanged to within a small accuracy $\epsilon$:  
$\sum_\mu p_\mu \| \rho - \rho_{\ell\mu}\|\leq\epsilon$. 
For example, a useful POVM is the set of Gaussian
quasi-projections: 
$A_{\mu} = (1/(2\pi)^{1/4}\Delta^{1/2})
\int_{-\infty}^{\infty} e^{-(a-\mu)^2/4\Delta^2}
|a\rangle \langle a| da $
where the normalization is chosen so that ${\rm tr}
A^\dagger_\mu A_\mu = 1$ and $\int_{-\infty}^\infty
A^\dagger_\mu A_\mu d\mu = {\bf I}$ (here there is
no need for the auxiliary index $i$).
If we write the single-system density matrix in the $a$ basis
as $\sum_{aa'} \alpha_{aa'} |a\rangle\langle a'|$, then the
measurement corresponding to the $A_\mu$ determines the value
of $\bar a= {\rm tr} \rho A$ to an accuracy $\Delta$, where
$A = \int a|a\rangle\langle a| da$.  In addition, 
the measurement has the effect of reducing the 
off-diagonal terms
of $\rho$ by an factor $e^{-(a-a')^2/2\Delta^2}$, corresponding
to a perturbation of size $\epsilon \approx \Delta A^2/2\Delta^2$,
where $\Delta A = \sqrt{{\rm tr} \rho A^2 - {\bar a}^2}$. 
If $\Delta >> \Delta A$ the measurement perturbs the system only weakly.
Of course, the more weakly the measurement perturbs the
system, the less information it acquires.
By making a weak measurement on all the systems in the set
simultaneously, however, one can obtain very precise
information about the single-system density matrix
while perturbing it only slightly.  Consider the
$N$-system POVM given by
$$A_{N\mu} = (1/(2\pi)^{1/4}\Delta^{1/2})^N
\int_{-\infty}^{\infty} e^{-(\sum_\ell a_\ell-N\mu)^2/4\Delta^2}
|a_1\rangle \langle a_1| \ldots |a_N\rangle \langle a_N| 
da_1\ldots da_N.$$
If a collective measurement corresponding to this POVM
is performed on the systems
in the set, one obtains the value of $\bar a$ to an accuracy
$\sqrt{ \Delta^2/N^2 + {\Delta A}^2/N}$, while
still perturbing the single-system density matrix by
the amount $\epsilon \approx \Delta A^2/2\Delta^2$. 
It can be clearly seen that in the limit $N\rightarrow\infty$
we can take $\Delta \propto \sqrt N$, giving an arbitrarily
accurate determination of $\bar a$ together with an arbitrarily 
small perturbation of the single-system density matrix.
After the measurement, the over-all density matrix is in
the form $\rho\otimes\ldots \otimes \rho + O(\epsilon)$, so that
the assumption of no correlation between the systems is
only true to order $\epsilon$.  In the limit $N\rightarrow\infty$,
$\epsilon\rightarrow 0$, however, the no-correlation assumption
still holds. 

Now we construct a model of how such a weak measurement 
might be performed.  Our model is analogous to weak 
measurements in NMR, in which each system in the set is
coupled weakly to the electromagnetic field in the 
measurement coil.  Couple each system to
the measurement apparatus via a single
continuous quantum variable (`pointer position'), described 
by an operator $Q=\int q |q\rangle\langle q| dq$, 
via a Hamiltonian coupling $\gamma A P$, 
where $P$ is the momentum corresponding to $Q$; $[P,Q]=i$.
This gives a dynamics for the system and pointer: $|a\rangle|q\rangle 
\rightarrow |a\rangle|q+a\gamma t\rangle$ over time $t$.
Now suppose that all $N$ systems are coupled symmetrically
to the pointer by an interaction $(A_1+\ldots+A_N)P$.
If the systems are all originally in the state $\rho_i=\rho$ as
above, and the pointer is originally in the 
state $|\psi\rangle = \int \psi(q) |q\rangle dq$, then the 
interaction between the systems and the measurement apparatus gives 
$$\eqalign{&\rho_1\otimes \ldots \otimes \rho_N\otimes 
|\psi\rangle\langle\psi|
 \rightarrow \rho_{SM}(t)\cr
&=\sum_{a_1a'_1\ldots a_Na'_N} \int dq dq' \psi(q) \bar\psi(q')
\rho_{a_1a'_1}\ldots \rho_{a_Na'_N}\cr
&|a_1\rangle\langle a'_1| \otimes\ldots \otimes 
|a_N\rangle\langle a'_N| 
 \otimes
|q+(a_1+\ldots+a_N)\gamma t \rangle
\langle q'+(a'_1+\ldots+a'_N)\gamma t|. \cr}\eqno(3)$$
One can then find
the state of the apparatus at time $t$ by taking 
$$\eqalign{\rho_M(t)=& {\rm tr}_S~ \rho_{SM}(t)  \cr 
=&\sum_{a_1\ldots a_N} \int dq dq' \psi(q) \bar\psi(q')
\rho_{a_1a_1}\ldots \rho_{a_Na_N} \cr
& |q+(a_1+\ldots+a_N)\gamma t \rangle
\langle q'+(a_1+\ldots+a_N)\gamma t|. \cr}\eqno(4)$$
That is, after the measurement the pointer registers
the sum of $N$ independent samples of $A$, where each result $a$ 
occurs with probability $p_a=\rho_{aa}$.
The sum is registered to 
an accuracy $\Delta Q = \sqrt{ \langle\psi|Q^2|\psi\rangle
- \langle\psi|Q|\psi\rangle^2}$. $\Delta Q$ measures the
initial spread of the pointer wave function $|\psi\rangle$. 
Accordingly, after the coupling of the pointer to the systems,
the pointer registers the result $\langle A \rangle = {\rm tr} \rho A
\pm \Delta A\sqrt{ 1/\epsilon N^2 + 1/N}$,
where $\epsilon = (\gamma t/\Delta Q)^2$ will be seen to be a measure of 
the degree of perturbation of each individual system,  and
$\Delta A =\sqrt{ {\rm tr} \rho A^2 -  ({\rm tr}\rho A)^2}$ 
is the standard deviation of $A$.

Now determine the amount of disturbance induced on the systems
in the set.  The state of any one of the systems in the set
after the coupling with the pointer is given by
tracing over all the other systems and the pointer state.
Since the systems are identical by symmetry, look just at
the first:
$$\eqalign{ \rho(t) =
&\sum_{aa'}\sum_{a_2\ldots a_N} \int dq dq' \psi(q) \bar\psi(q')
\rho_{aa'} |a\rangle\langle a'|
\rho_{a_2a_2}\ldots \rho_{a_Na_N}\cr
&\langle q'+(a'+ a_2+\ldots+a_N)\gamma t 
|q+(a+ a_2+\ldots+a_N)\gamma t \rangle.\cr
=&\sum_{aa'} \int dq  \psi(q) \bar\psi(q+(a-a')\gamma t)
\rho_{aa'} |a\rangle\langle a'|
\cr}\eqno(5)$$
That is, the off-diagonal parts of $\rho_1$ are
reduced by an amount $1-\int dq  \psi(q) 
\bar\psi(q+(a-a')\gamma t)$.  A convenient initial pointer state
$|\psi\rangle$ is a Gaussian wave packet centered
at 0 with standard deviation $\Delta Q$ (analogous
to a coherent state of the electromagnetic field).  
In this case, it is easily seen that the effect of the 
coupling to the pointer is to multiply the $aa'$ off-diagonal
terms of $\rho$ by a factor $e^{- (\gamma t (a-a'))^2 /2\Delta Q^2} $.
That is, when $(\gamma t\Delta A)^2/2\Delta Q^2 \approx \epsilon << 1$,
the effect of coupling each member of the set
to the same pointer is essentially the same as the effect
of coupling each member to a different measuring apparatus,
with a perturbation of size $\epsilon  = (\gamma t \Delta A/\Delta Q)^2$.
This model of measurement can be seen to be equivalent
to the abstract POVM given above.

It is interesting to note that
the `weakness' of this model of measurement can be tuned by adjusting 
the spread $\Delta Q$ of the initial pointer
wave packet.  As $\Delta Q$ becomes small, the measurement becomes
stronger and stronger, revealing more information about an individual
system while perturbing its wave function more and more.  In the
limit that $\Delta Q\rightarrow 0$, this model reduces to von
Neumann's original model of strong measurement.  $\Delta Q$
acts as a knob that allows us to tune continuously from weak
to strong measurement.  

We can weakly measure several observables ${\cal O}_\ell$ 
simultaneously by adjoining several pointer variables $Q_\ell$ 
and coupling $\sum_\ell \gamma_\ell{\cal O}_\ell Q_\ell$.
In the limit $\gamma\rightarrow 0$, $N\rightarrow\infty$,
the $\ell$'th pointer provides an accurate assessment
of $\langle O_\ell \rangle$ while perturbing each system
by as small an amount as desired.  Note that ${\cal O}_\ell$
need not commute with each other: in the weak measurement limit 
where the $\Delta Q_\ell$ are large and $\gamma_\ell$ are
small, the measurements do not interfere with eachother.  
By monitoring $d^2-1$ observables, one can obtain an assessment 
of all terms in the density matrix simultaneously. 

This concludes the detailed discussion of weak measurement.
To summarize: by adjoining a suitable measuring apparatus 
and making the number of systems $N$ in the set large, one
can obtain the density matrix to a precision 
$\delta=\sqrt{1/\epsilon N^2+1/N}$ while perturbing each system 
by an amount $\epsilon (d^2-1)$.  By making $N$ sufficiently large, 
the single-system reduced density matrix can be determined to
arbitrary precision while perturbing each system by an
arbitrarily small amount.  It can be seen that
the detailed model gives the same results as the abstract
model of weak measurement given above.  

It is interesting to investigate whether the underlying statistics (Fermionic
or Bosonic) of the systems in the set affect the results above.
Since both wave functions and interactions
are assumed to be symmetric, the results derived above hold
equally well for Bosonic systems.  If the systems are Fermionic,
in contrast, they cannot be in completely identical states.  However,
if each system possesses additional degrees of freedom
(position, for example, in the case of nuclear spins) that
do not figure in the interaction with the measuring apparatus,
then the discussion above applies to fermions as well.

In fact, although we have assumed a
symmetric situation in which the systems are described by 
identical density matrices $\rho$, this restriction is not
necessary.  If the systems are prepared in the uncorrelated state
$\rho_1 \otimes \ldots \otimes \rho_N$, where in general $\rho_i\neq\rho_j$,
then the entire set of results derived
here applies to the determination and control of
the {\it average} single-system density matrix
$\bar\rho = (1/N) \sum_i \rho_i$.  

Let us now assume that we can perform arbitrary weak measurements
on a set of quantum systems, and feed the results of
those measurements back continuously and coherently using
the well-known techniques of coherent control.  That is,
assume that we can implement arbitrary nonlinear unitary
transformations as in equation (1b) and nonlinear Schr\"odinger
equations as in equation (2).  How might this technique be applied?
 
The first potential use of this technique is simulation: a weak 
feedback controller could be used to simulate the dynamics of 
a variety of systems that obey a nonlinear Schr\"odinger equation.  
Nonlinear Schr\"odinger equations tend to arise in sets
of weakly coupled quantum systems: as noted above in the context
of the nonlinear Bloch equation, such coupled sets can be
thought of as naturally occurring examples of weak feedback.
For example, weak feedback can be used to simulate any set 
of systems that can be adequately described by a mean field theory, 
in which each system is coupled weakly to the expectation value
of some operators on the ensemble as a whole. 

The use of a nonlinear Schr\"odinger equation is common in 
quantum optics to describe the evolution of photons that are 
weakly interacting with matter, as in a nonlinear
optical fiber (23-24).  As just noted, such an effect can be thought of as
a naturally occurring example of quantum feedback by weak 
measurements: the atoms in the fiber weakly monitor and act on 
the photons.  The use of weak feedback to create nonlinearities has the
advantage that the form and strength of the nonlinearity
induced by the quantum controller can be varied at will.
For example, an optical weak feedback apparatus could be constructed
by instrumenting a fiber with photodetectors and feeding
back their signals to the fiber via electro-optic modulators (25).
Such an optical controller could be used as 
a quantum analog computer to investigate the effect of time and 
spatially varying nonlinearities on the propagation of light
down the fiber.  It is important to note that such a fiber need 
not itself be nonlinear: all the nonlinearity could be supplied 
by the controller.  In addition, weak feedback could be used to 
create and investigate the properties of optical solitons in
a variety of nonlinear media.  

As noted by Haus (23), systems that obey nonlinear Schr\"odinger
equations can be used to create Schr\"odinger's cats---quantum
systems that exist in superpositions of two widely differing 
quasi-classical states.  Although optical fibers are lossy and tend
to introduce decoherence, single-mode optical cavities of the
sort constructed by Kimble {\it et al.} are good candidates
for control by weak feedback (26).  The very high Q of such cavities 
implies that the mode in the cavity is only weakly coupled to 
modes outside the cavity.  Heterodyne monitoring of the cavity
field therefore constitutes a weak measurement on the photons
in the cavity.  Nonlinearities induced by weak measurement could 
be used to create Schr\"odinger's cat states \`a la Haus.
Wiseman and Milburn have also proposed a cavity quantum electrodynamics
enactment of feedback via weak measurements to perform optical
squeezing (7).

Finally, quantum feedback via weak measurement could be used to
create a novel form of quantum chaos.  The usual linear Schr\"odinger
equation does not exhibit sensitive dependence to initial 
conditions: the `distance' between any two states $|\phi\rangle$ and
$|\phi'\rangle$, as measured by their inner product $\langle \phi|
\phi'\rangle$, remains constant (27-28).  (Traditionally quantum chaos
is not the study of sensitive dependence of quantum trajectories
on initial conditions, but rather the study of quantized versions 
of classical chaotic systems.)  The nonlinear Schr\"odinger equation,
in contrast, need not preserve distances between quantum states,
and {\it can} exhibit sensitive dependence on initial conditions (29-30).
Quantum feedback via weak measurement, because
it can be used to effect arbitrary nonlinear Schr\"odinger equations,
offers unique opportunities for investigating the sensitive 
dependence of quantum trajectories on initial conditions.
Such ensemble quantum chaos could be used for example to construct
a `Schr\"odinger microscope' to detect and
amplify small differences in quantum wave functions.

We close by examining the relationship between nonlinearity
induced by feedback of weak measurements and intrinsically
nonlinear quantum mechanics.  (Once again, the nonlinearity
discussed in this paper is an {\it effective} nonlinearity: the
underlying quantum dynamics of weak feedback is linear.)
Nonlinear Schr\"odinger equations
of the type found in equation (2) above are common in nonlinear
quantum mechanics (31).  Nonlinear quantum mechanics is
known to exhibit a number of pathologies, including superluminal
communication (32), violations of the second law of thermodynamics (33),
and the ability to solve hard computational problems (34).
Since nonlinearity induced by weak feedback occurs entirely
within the conventional framework of quantum mechanics, it
cannot exhibit the first two of these pathologies.  It might
allow the solution of hard computational problems by the
mechanism of the previous paragraph, viz., constructing a 
`Schr\"odinger microscope' to detect small perturbations in
the wave function of a quantum computer (see also (35)); however, to 
obtain an exponential speed-up over classical computations one needs
to amplify exponentially small differences in the wave function
which in turn requires an exponentially large number of systems
in the set.  Nonetheless, it may be the case that weak feedback
can be used to provide polynomial speed-ups to computational
problems.

\vfill
\noindent{\it Acknowledgements:} This work was supported by the ARO
and by DARPA under the Quantum Information and Computation (QUIC)
initiative.
\vfil\eject

\centerline{\bf References}

\noindent(1) G.M. Huang, T.J. Tarn, J.W. Clark, On the
controllability of quantum-mechanical systems.
{\it J. Math. Phys.} {\bf 24}(11), 2608-2618 (1983).

\noindent(2) R.W. Brockett, R.S. Millman, H.J. Sussman,
eds., {\it Differential Geometric Control Theory}
(Birkhauser, Boston, 1983).  Z. Li, J.F. Canney, eds., {\it
Nonholonomic Motion Planning} (Kluwer Academic, Boston,
1993).

\noindent(3) A. Blaquiere, S. Diner, G. Lochak, eds.,
Information, Complexity and Control in Quantum Physics.
(Springer-Verlag, New York, 1987).
A. Blaquiere,  Modeling and Control of
Systems in Engineering, Quantum Mechanics, Economics and
Biosciences. (Springer-Verlag, New York, 1989).

\noindent(4) A.G. Butkovskiy, Yu.I. Samoilenko, Control
of Quantum-Mechanical Processes and Systems. (Kluwer
Academic, Dordrecht, 1990).

\noindent(5) H. Ezawa, Y. Murayama, eds., Quantum
Control and Measurement. (North-Holland, Amsterdam, 1993).

\noindent(6) A. Peirce, M. Dahleh, H. Rabitz, Optimal
Control of Quantum-Mechanical Systems: Existence, Numerical
Approximation, and Applications, {\it Phys. Rev. A} {\bf 37},
4950-4964 (1988); M. Dahleh, A.P. Peirce, H. Rabitz,
Optimal Control of Uncertain Quantum Systems, {\it Phys. Rev. A}
{\bf 42}, 1065-1079 (1990); R.S. Judson, H.
Rabitz, Teaching Lasers to Control Molecules,
{\it Phys. Rev. Lett.} {\bf 68}, 1500-1503 (1992); W.S.
Warren, H. Rabitz, M. Dahleh, Coherent Control of Quantum
Dynamics: The Dream is Alive, {\it Science} {\bf 259}, 1581-1589
(1993); V. Ramakrishna, M.V. Salapaka, M. Dahleh, H. Rabitz,
A. Peirce, Controllability of Molecular Systems,
{\it Phys. Rev. A} {\bf 51}, 960-966 (1995).

\noindent(7) H.M. Wiseman, G.J. Milburn, Quantum Theory
of Optical Feedback via Homodyne Detection, {\it Phys. Rev. Lett.}
{\bf 70}, 548-551 (1993); H.M. Wiseman, G.J. Milburn, Squeezing
vi feedback, {\it Phys. Rev. A} {\bf 49}, 1350-1366 (1994);
H.M. Wiseman, Quantum Theory of Continuous Feedback,
{\it Phys. Rev. A} {\bf 49}, 2133-2150 (1994).

\noindent(8) A. Peres, Zeno Paradox in Quantum Theory,
{\it Am. J. Phys.} {\bf 48}, 931-932 (1980).

\noindent(9) V.B. Braginsky, Y.I. Yorontsov, K.S. Thorne,
Quantum Nondemolition Measurements, {\it Science} {\bf 209},
547-557 (1980).

\noindent(10) C.M. Caves, K.S. Thorne, R.W.P. Drever, V.D.
Sandberg, M. Zimmerman, On the Measurement of a Weak Classical
Force Coupled to a Quantum Mechanical Oscillator,
{\it Rev. Mod. Phys.} {\bf 52}, 341-392 (1980).

\noindent(11) S. Lloyd, Controllability and Observability of
Hamiltonian Quantum Systems, submitted to {\it Physical Review
A}.

\noindent(12) S. Lloyd, R.J. Nelson, Y. Weinstein, D. Cory,
Experimental Demonstration of Coherent Quantum Feedback,
submitted to {\it Physical Review Letters}.

\noindent(13) O.R. Ernst, G. Bodenhausen, A. Wokaun,
Principles of Nuclear Magnetic Resonance in One and Two
Dimensions.  (Oxford University Press, Oxford, 1987).

\noindent(14) Y. Aharonov, D.Z. Albert, L. Vaidman,
{\it Phys. Rev. Lett.} {\bf 60}, 1351-1354 (1988).
Y. Aharonov, L. Vaidman,
{\it Phys. Rev. Lett.} {\bf 62}, 2327 (1989).

\noindent(15) A.J. Leggett, {\it Phys. Rev. Lett.} {\bf 62}, 
2325 (1989).

\noindent(16) A. Peres, {\it Phys. Rev. Lett.} {\bf 62}, 
2326 (1989).

\noindent(17) P. Broekart, J. Jeener, {\it J. Magn. Reson. Ser. A}
{\bf 113}, 60 (1995).

\noindent(18) A. Louis-Joseph, Y.-Y. Lallemand, {\it J. Biomol.
NMR} {\bf 5}(2), 212-216 (1995).

\noindent(19) S. Vathyam, S. Lee, W.S. Warren, {\it Science} {\bf
272}, 92-96 (5 April 1996).  W. Richter, S. Lee, W.S. Warren,
Q. He, {\it Science} {\bf 276}, 654-657 (3 February 1997).

\noindent(20) K. Kraus, {\it States, Effects, and Operations:
Fundamental Notions of Quantum Theory}, (Springer-Verlag, Berlin, 1983).

\noindent(21) R. Alicki, K. Lendi, {\it Quantum Dynamical Semigroups
and Applications}, (Springer-Verlag, Berlin, 1987).

\noindent(22) J. von Neumann, {\it Mathematische Grundlagen
der Quantenmechanik}, (Springer, Berlin, 1932).

\noindent(23) H.A. Haus, `Nonlinear Optics,' in {\it Waveguide
Optoelectronics,} J.H. Marsh and R.M. De La Rue, eds.,
ch. 11, pp. 225-288 (Kluwer Academic, Netherlands, 1992).
Y. Lai, H. Haus, {\it Phys. Rev. A} {\bf 40}, 844-853,
854-866 (1989).  A. Mecozzi, J.D. Moores, H.A. Haus, Y. Lai,
{\it J. Opt. Soc. Am. B} {\bf 9}, 1350-1357 (1992).
H. Haus, F.X. Kartner, {\it Phys. Rev. A} {\bf 46},
R1175-R1176 (1992) 

\noindent(24) D.C. Calvo, T.R. Akylas {\it Phys. Rev. E} {\bf 56},
4757-4764 (1997).

\noindent(25) D.J. Jones, K.L. Hall, H.A. Haus, E.P. Ippen,
{\it Opt. Lett.} {\bf 23}, 177-179 (1998).

\noindent(26) H. Mabuchi and H.J. Kimble, {\it Optics Letters} {\bf
19}, 749 (1993).
A.B. Matsko, S.P. Vyatchanin, H. Mabuchi, H.J. Kimble,
{\it Physics Letters A} {\bf 192}, 175 (1994).
H.J. Kimble, O. Carnal, N. Georgiades, H. Mabuchi, E.S.
Polzik, R.J. Thompson, Q.A. Turchette, in {\it Proc. Int. Conf. on
Atomic Physics.}
A.S. Parkins, P. Marte, P. Zoller, H.J. Kimble, {\it Phys. Rev.
Lett.} {\bf 71}, 3095 (1993).

\noindent(27) E.R. Pike, S. Sarkar, {\it Quantum Measurement and
Chaos}, Nato ASI Series B161, (Plenum, New York 1986).

\noindent(28) P. Cvitanovi\'c, I. Percival, A. Wirzba,
{\it Quantum Chaos--Quantum Measurement}, Nato ASI Series C358,
(Kluwer, Dordrecht, 1992).

\noindent(29) C.P. Smith, R. Dykstra, {\it Opt. Comm.} {\bf 126},
69-74 (1996).

\noindent(30) J.-Y. Liu, Z.-B. Wu, W.-W. Zheng, {\it Commun. Theor.
Phys.} {\bf 25}, 149-158 (1996).

\noindent(31) Detailed discussions of nonlinear quantum mechanics
can be found in M. Czachor, {\it Found. Phys. Lett.} {\bf 4}, 351 (1991); 
{\it Phys. Rev. A} {\bf 53}, 1310-1315 (1996);
{\it Phys. Rev. A} {\bf 57}, 4122 (1998); {\it Phys. Lett. A} {\bf 225},
1-12 (1997); {\it Int. J. Theor. Phys.} {\bf 38}, 475 (1999);  
{\it Acta. Phys. Slov.} {\bf 48}, 157 (1998);
M. Czachor and M. Marciniak, {\it Phys. Lett. A} {\bf 239}, 353 (1998);
M. Czachor and M. Kuna, {\it Phys. Rev. A} {\bf 58}, 128 (1998);
S.B. Leble and M. Czachor, {\it Phys. Rev. E} {\bf 58}, 7091 (1998);
M. Kuna, M. Czachor, and S.B.Leble, {\it Phys. Lett. A} {\bf 255}, 42 (1999);
M. Czachor and J. Naudts, {\it Phys. Rev. E} {\bf 59}, R2497 (1999);
M. Czachor, M. Kuna, S.B. Leble, and J. Naudts, quant-ph/9904110.

\noindent(32) S. Weinberg, {\it Phys. Rev. Lett.} {\bf 62},
485 (1989).  N. Gisin, {\it Phys. Lett. A} {\bf 113}, 1 (1990).
J. Polchinski, {\it Phys. Rev. Lett.} {\bf 66}, 397 (1991).

\noindent(33) A. Peres, {\it Phys. Rev. Lett.} {\bf 63}, 1114 (1989).

\noindent(34) D. Abrams, S. Lloyd, {\it Phys. Rev. Lett.} {bf  81},
3992-3995 (1998).

\noindent(35) D.G. Cory, A.F. Fahmy, T.F. Havel, Nuclear Magnetic
Resonance Spectroscopy: an experimentally accessible paradigm
for quantum computing, in {\it PhysComp96}, Proceedings of the
Fourth Workshop on Physics and Computation, T. Toffoli, M. Biafore,
J. Le\~ao, eds., New England Complex Systems Institute, 1996,
pp. 87-91.

\vfill\eject\end